\documentclass[aps,prd,twocolumn,showpacs,amsmath,amssymb]{revtex4-1}
\usepackage{amsmath}
\usepackage{graphicx}
\usepackage{subfigure}
\usepackage{epstopdf}
\usepackage{color}
\usepackage{multirow}
\usepackage{setspace}
\usepackage{overpic}
\usepackage{amssymb}
\usepackage{lineno}
\usepackage{bm}
\usepackage{rotating}
\usepackage{makecell}
\usepackage[utf8]{inputenc}
\usepackage{hyperref}

\usepackage{lineno}
\usepackage{setspace}

\hypersetup{
	colorlinks=true,
	linkcolor=blue,
	filecolor=blue,
	urlcolor=blue,
	citecolor=blue,
}
\let\oldequation\equation
\let\oldendequation\endequation

\renewenvironment{equation}
  {\linenomathNonumbers\oldequation}
  {\oldendequation\endlinenomath}

\newcommand{\BESIIIorcid}[1]{\href{https://orcid.org/#1}{\hspace*{0.1em}\raisebox{-0.45ex}{\includegraphics[width=1em]{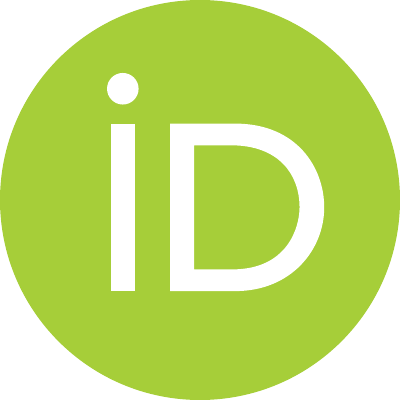}}}}

\begin{document}


\title{\bf \boldmath
Search for the rare decays of $D\to h(h^{(')})e^{+}e^{-}$}

\author{
M.~Ablikim$^{1}$\BESIIIorcid{0000-0002-3935-619X},
M.~N.~Achasov$^{4,d}$\BESIIIorcid{0000-0002-9400-8622},
P.~Adlarson$^{82}$\BESIIIorcid{0000-0001-6280-3851},
X.~C.~Ai$^{88}$\BESIIIorcid{0000-0003-3856-2415},
C.~S.~Akondi$^{31A,31B}$\BESIIIorcid{0000-0001-6303-5217},
R.~Aliberti$^{39}$\BESIIIorcid{0000-0003-3500-4012},
A.~Amoroso$^{81A,81C}$\BESIIIorcid{0000-0002-3095-8610},
Q.~An$^{78,64,\dagger}$,
Y.~H.~An$^{88}$\BESIIIorcid{0009-0008-3419-0849},
Y.~Bai$^{62}$\BESIIIorcid{0000-0001-6593-5665}
% ...
\\
\vspace{0.2cm}
(BESIII Collaboration)\\
\vspace{0.2cm} {\it
$^{1}$ Institute of High Energy Physics, Beijing 100049, People's Republic of China\\
% ...
$^{4}$ Budker Institute of Nuclear Physics SB RAS (BINP), Novosibirsk 630090, Russia\\
% ...
$^{31}$ INFN Sezione di Ferrara, (A)INFN Sezione di Ferrara, I-44122, Ferrara, Italy; (B)University of Ferrara, I-44122, Ferrara, Italy\\
% ...
$^{39}$ Johannes Gutenberg University of Mainz, Johann-Joachim-Becher-Weg 45, D-55099 Mainz, Germany\\
% ...
$^{62}$ Southeast University, Nanjing 211100, People's Republic of China\\
% ...
$^{64}$ State Key Laboratory of Particle Detection and Electronics, Beijing 100049, Hefei 230026, People's Republic of China\\
% ...
$^{78}$ University of Science and Technology of China, Hefei 230026, People's Republic of China\\
% ...
$^{81}$ University of Turin and INFN, (A)University of Turin, I-10125, Turin, Italy; (B)University of Eastern Piedmont, I-15121, Alessandria, Italy; (C)INFN, I-10125, Turin, Italy\\
$^{82}$ Uppsala University, Box 516, SE-75120 Uppsala, Sweden\\
% ...
$^{88}$ Zhengzhou University, Zhengzhou 450001, People's Republic of China\\
\vspace{0.2cm}
$^{\dagger}$ Deceased\\
% ...
$^{d}$ Also at the Novosibirsk State University, Novosibirsk, 630090, Russia\\
% ...
}
}

\begin{abstract}
We search for 15 rare decays of $D$ mesons to hadrons accompanied by an electron-positron pair $D\to h(h^{(')})e^{+}e^{-}$, based on 20.3 fb$^{-1}$ of $e^+ e^-$ collision 
data collected at the center-of-mass energy of 3.773 GeV with the BESIII detector at BEPCII.
No significant signals are observed, and the corresponding upper limits on the branching fractions at the 90\% confidence level are determined.
The sensitivities of the results are at the level of $10^{-6}$ $\sim$ $10^{-7}$.
The upper limits on the branching fractions for the $D^+\to \rho^{+}  e^+ e^-$, $D^+\to K^{*+}  e^+ e^-$, $D^0\to K_S^0 K_S^0 e^+ e^-$, $D^0\to \pi^0 \pi^0 e^+ e^-$ and 
$D^0\to \eta^{\prime}  e^+ e^-$ decay channels are measured for the first time.
For the $D^0\to \pi^0 e^+ e^-$, $D^0\to \eta e^+ e^-$, $D^0\to \omega e^+ e^-$, $D^0\to K_S^0 e^+ e^-$, $D^+\to \pi^+ \pi^0 e^+ e^-$, $D^+\to K^+ \pi^0 e^+ e^-$, $D^+\to 
\pi^+ K_S^0 e^+ e^-$ and $D^+\to K^+ K_S^0 e^+ e^-$ decay channels, the upper limits on the branching fractions are determined, with an improvement of at least a factor of 
four compared to previous searches.
The upper limits on the branching fractions for the $D^0\to \rho^{0} e^+ e^-$ and $D^0\to \phi e^+ e^-$ decay channels are set at $0.7 \times 10^{-6}$ and $4.6 \times 
10^{-6}$, respectively.

\end{abstract}

\maketitle

\oddsidemargin  -0.2cm
\evensidemargin -0.2cm

\section{Introduction}

In the Standard Model (SM), the flavor-changing-neutral-current (FCNC) processes are strongly  suppressed by the Glashow-Iliopoulos-Maiani (GIM) mechanism~\cite{GIM}  and can only occur at loop level. Such processes have been observed in $K$ and $B$ meson decays, {\it e.g.}, the decays $K^{\pm} \to 
\pi^{\pm}\mu^{+}\mu^{-}$~\cite{ref_K}, and $B^{0} \to K^{*0}\mu^{+}\mu^{-}$~\cite{ref_B}, where heavy virtual quarks, particularly the top quark, contribute significantly in the loops.
By contrast, in the charm sector there is no heavy particle like the top quark in the loop diagram $c \to ul^+l^-$ ($l$ = $e$ or $\mu$). Consequently, the GIM suppression in charm FCNC decays is much stronger than that in the corresponding processes of $B$ and $K$ mesons, leading to a theoretically very small branching fraction (BF) that would not exceed the level of $10^{-9}$~\cite{CtoU, cxll, hhll}. 

On the other side, possible new physics (NP) can significantly increase the decay rates of these short-distance (SD) processes. Hence, they can serve as clean channels in experiments to search for NP~\cite{CtoU, cxll}.
However, these decay rates of $D$ mesons also receive substantial contribution from long distance (LD) effects through vector meson ($V$) decays, like $D \to h V,V \to l^+l^-$, where $h$ is a hadron.
These LD effects can mimic the signal and typically yield branching fractions of order 
$10^{-6}$~\cite{cxll, hhll}, thereby overshadowing the SD component.
To disentangle the SD effects, a measurement of the angular dependence or CP asymmetry~\cite{hhll} is essential. Moreover, comparing the electron and muon channels in these decays provides a clean test of lepton flavor universality, another key prediction of the SM~\cite{hhll,hhll1}.

Comprehensive experimental measurements of other three- and four-body decays to different hadronic final states are still of great interest.
Theoretically, potential NP contributions can interfere with the hadronic interactions in various forms and magnitudes thereby affecting the sensitivity to relevant NP effects~\cite{hhll2,hhll3}.
Recently, the LHCb Collaboration reported the latest results on the search for $D^0$ meson decays to $\pi^{+}\pi^{-}e^{+}e^{-}$ and $K^{+}K^{-}e^{+}e^{-}$ final states~\cite{LHCb}. The results are consistent with the BFs of the $D^0 \to \pi^+\pi^- \mu^+\mu^-$ and $D^0 \to K^+K^- \mu^+\mu^-$ decays~\cite{LHCb1} measured by the LHCb Collaboration, thereby verifying lepton flavor universality.
The decay $D^0 \to \pi^{+}\pi^{-}e^{+}e^{-}$ has been observed for the first time where the $e^{+}e^{-}$ pair originates from the decay of a $\phi$ or $\rho^0/\omega$ meson~\cite{LHCb}. While, no evidence is found for the decay $D^0 \to K^{+}K^{-}e^{+}e^{-}$~\cite{LHCb}.
The same long-distance resonant contributions $\rho^0/\omega\to e^{+}e^{-}$ have been observed in the decay $D^0 \to K^{-}\pi^{+}e^{+}e^{-}$ by \textit{BABAR} and Belle II collaboration~\cite{BaBar,Bell}. 
Based on the 2.93 fb$^{-1}$ dataset collected at $\sqrt{s} = 3.773~\text{GeV}$, BESIII collaboration has performed a study of the \mbox{$D \to h(h^{(')})e^{+}e^{-}$} processes, benefiting from the clean sample of charm-meson decays and adopting the double-tag (DT) method~\cite{mark3}, where $h^{(')}$ are hadrons.
The most stringent upper limits have been set on the BFs for those involving neutral $\pi^0$, $\eta$ and $K_S^0$ mesons in the final state~\cite{sys_hhee}.
More precise measurements for these $D \to h(h^{(')})e^{+}e^{-}$ decays based on the larger dataset at BESIII would help to further drive theoretical progress and bridge the current gaps in detailed theoretical calculations.

In this paper, with an integrated luminosity of 20.3 fb$^{-1}$~\cite{Ablikim:2019hff} collected with the BESIII detector at 
$\sqrt{s} = 3.773~\text{GeV}$, we search for 15 rare decays of $D\to h(h^{(')})e^{+}e^{-}$.
To reduce possible bias, a blind analysis is carried out based on Monte Carlo (MC) simulations to validate the analysis strategy, the results are obtained only after the analysis strategy is fixed.

\section{BESIII detector and Monte Carlo simulation}
The BESIII detector~\cite{Ablikim:2009aa} records symmetric $e^+e^-$ collisions
provided by the BEPCII storage ring~\cite{Yu:IPAC2016-TUYA01}
in the center-of-mass energy range from 1.84 to  4.95 GeV,
with a peak luminosity of $1.1 \times 10^{33}\;\text{cm}^{-2}\text{s}^{-1}$
achieved at $\sqrt{s} = 3.773\;\text{GeV}$.
BESIII has collected large data samples in this energy region~\cite{Ablikim:2019hff,EcmsMea,EventFilter,You_sun}.
The cylindrical core of the BESIII detector covers 93\% of the full solid angle and consists of a helium-based multilayer drift chamber~(MDC), a time-of-flight
system~(TOF), and a CsI(Tl) electromagnetic calorimeter~(EMC), which are all enclosed in a superconducting solenoidal magnet providing a 1.0~T magnetic field.
The solenoid is supported by an octagonal flux-return yoke with resistive plate counter muon identification modules interleaved with steel.
The charged-particle momentum resolution at $1~{\rm GeV}/c$ is $0.5\%$, and the ${\rm d}E/{\rm d}x$ resolution is $6\%$ for electrons from Bhabha scattering.
The EMC measures photon energies with a
resolution of $2.5\%$ ($5\%$) at $1$~GeV in the barrel (end cap) region.
The time resolution in the plastic scintillator TOF barrel region is 68~ps, while that in the end cap region was 110~ps.
The end cap TOF system was upgraded in 2015 using multigap resistive plate chamber technology, providing a time resolution of 60~ps, which benefits 85\% of the data used in 
this analysis~\cite{etof}.

Simulated event samples produced with the {\sc geant4}-based~\cite{geant4} MC package which includes the geometric description of the BESIII detector and the
detector response, are used to determine the detection efficiency and estimate the backgrounds.
The simulation includes the beam-energy spread and initial-state radiation (ISR) in the $e^+e^-$
annihilations modeled with the generator {\sc kkmc}~\cite{kkmc}.
The inclusive MC sample includes the production of $D\bar{D}$ pairs (including quantum coherence for the neutral $D$ channels), the non-$D\bar{D}$ decays of the $\psi(3770)$, 
the ISR production of the $J/\psi$ and $\psi(3686)$ states, and the continuum processes incorporated in {\sc kkmc}~\cite{kkmc}.
All particle decays are modelled with {\sc evtgen}~\cite{evtgen} using the BFs either taken from the Particle Data Group~\cite{pdg2025}, when available, or otherwise estimated 
with {\sc lundcharm}~\cite{lundcharm}.
Final-state radiation (FSR) from charged particles is incorporated with the {\sc photos} package~\cite{photos}.
Signal MC simulation is utilized for the study of detection efficiency. The signal simulation incorporates both the primary LD contribution from the $h(h^{(\prime)}) \omega, 
\omega \to e^+ e^-$ process and the SD contribution described by a phase space (PHSP) model across different $e^{+}e^{-}$ mass regions.

\section{Method}

The DT method is used in this work~\cite{mark3}. At $\sqrt s=3.773$\rm \,GeV, the $D^0\bar D^0$ or $D^+D^-$ meson pairs are produced from $\psi(3770)$ decays without 
accompanying hadrons, which provides an ideal opportunity to study rare decays of $D$ mesons.
In the first step, the single-tag (ST) $\bar D^0$ mesons are reconstructed via the hadronic-decay modes of
$\bar D^0\to K^+\pi^-$, $K^+\pi^-\pi^0$, and $K^+\pi^-\pi^-\pi^+$;
while the ST $D^-$ mesons are reconstructed via the decays
$D^-\to K^{+}\pi^{-}\pi^{-}$, $K^0_{S}\pi^{-}$, $K^{+}\pi^{-}\pi^{-}\pi^{0}$, $K^0_{S}\pi^{-}\pi^{0}$,
$K^0_{S}\pi^{+}\pi^{-}\pi^{-}$, and $K^{+}K^{-}\pi^{-}$.
Then the signal $D$ candidates are reconstructed with the remaining tracks which have not been used in the ST selection.
The event, in which the rare decay $D\to h(h^{(')})e^{+}e^{-}$ is reconstructed in the system recoiling against the ST $\bar D$ meson, is called a DT event~\cite{mark3}.
The BFs of the $D \to h(h^{(')})e^{+}e^{-}$ decays are determined by
\begin{equation}
    \label{br}
   {
\mathcal{B}=\frac{n_{\mathrm{sig,tag}}}{\sum_{i}n_{\mathrm{tag}}^{i}\cdot\frac{\varepsilon_{\mathrm{sig,tag}}^{i}}{\varepsilon_{\mathrm{tag}}^{i}}}.}
    \end{equation}
Here, $i$ denotes the different ST modes of hadronic decays, and $n_\mathrm{tag}^{i}$ is the yield of the $\bar{D}$ meson of ST tag mode $i$. $n_\mathrm{sig, tag}$ is the 
number of $D$ rare decay signal events in which an ST $D$ meson is detected. Finally, $\varepsilon_{\mathrm{tag}}^i$ and $\varepsilon_{\mathrm{sig, tag}}^i$ are the 
corresponding ST and DT detection efficiencies. Note that in this paper, charge conjugated modes are always implied.

\section{Single Tag selection}

Charged tracks detected in the MDC (except for those used for $K^0_S$ reconstruction) are required to be within a polar angle ($\theta$) range of $|\rm{cos\theta}|<0.93$, 
where $\theta$ is defined with respect to the $z$-axis, which is the symmetry axis of the MDC.
The distance of closest approach to the interaction point~(IP) along the $z$-axis, $|V_{z}|$, must be less than 10 cm, and in the transverse plane, $|V_{xy}|$, less than 1 cm.
Charged kaons and pions are identified using likelihoods $\mathcal{L}(h)~(h=K,\pi)$ constructed from measurements of specific ionization energy loss (d$E$/d$x$) in the MDC and flight time in the TOF. A track is assigned as a kaon if $\mathcal{L}(K) > \mathcal{L}(\pi)$, and as a pion if $\mathcal{L}(\pi) > \mathcal{L}(K)$.

Each $K_{S}^0$ candidate is reconstructed from two oppositely charged tracks satisfying $|V_{z}|<20$~cm.
The two charged tracks are assigned as $\pi^+\pi^-$ without imposing PID criteria.
They are constrained to originate from a common vertex, requiring an invariant mass within $(0.487,0.511)$~GeV/$c^2$.
The decay length of the $K_S^0$ candidate is required to be separated from the IP by more than twice the vertex resolution, which encompasses both the primary and secondary 
vertices.
The quality of the vertex fits (primary-vertex fit and secondary-vertex fit) is ensured by requiring $\chi^2 < 100$.

Photon candidates are identified using showers in the EMC. The $\pi^0$ candidates with both photons from the end cap are rejected because of poor resolution. The deposited 
energy of each shower must be more than 25~MeV in the barrel region ($|\!\cos \theta|< 0.80$) and more than 50~MeV in the end-cap region ($0.86 <|\!\cos \theta|< 0.92$).
Showers are required to be separated from charged tracks by an angle greater than $10^\circ$ in order to eliminate activity induced by charged particles.
To suppress electronic noise and showers unrelated to the event, the difference between the EMC time and the event start time is required to be within [0, 700]\,ns.
For $\pi^0$ candidates, the invariant mass of the photon pair is required to be within $(0.115,\,0.150)$\,GeV$/c^{2}$.
To improve the resolution, a kinematic fit is performed, where the diphoton invariant mass is constrained to the known $\pi^{0}$ mass~\cite{pdg2025}. The $\chi^{2}$ of the fit 
is required to be less than 50.
The momenta obtained from the kinematic fit are used in the subsequent analysis.

In the selection of $\bar D^0 \to K^+\pi^-$ events, cosmic rays and Bhabha backgrounds are suppressed by applying the same criteria described in Ref.~\cite{deltakpi}.
The two charged tracks are required to have a TOF time difference of less than 5~ns and must not be identified as a muon–antimuon or electron–positron pair.
Furthermore, each event must contain either at least one EMC shower with deposited energy exceeding 50~MeV, or at least one additional charged track detected in the MDC.

To separate the ST $\bar D$ mesons from combinatorial backgrounds, we define the energy difference $\Delta E\equiv E_{\bar D}-E_{\mathrm{beam}}$ and the beam-constrained mass 
$M_{\rm BC}\equiv\sqrt{E_{\mathrm{beam}}^{2}/c^{4}-|\vec{p}_{\bar D}|^{2}/c^{2}}$, where $E_{\mathrm{beam}}$ is the beam energy, and $E_{\bar D}$ and $\vec{p}_{\bar D}$ are 
the total energy and momentum of the $\bar D$ candidate in the $e^+e^-$ center-of-mass frame, respectively.
If there is more than one $\bar D$ candidate in a given ST mode, the candidate with the smallest value of $|\Delta E|$ is kept for the subsequent analysis. The $\Delta E$ 
requirements and ST efficiencies are listed in Table~\ref{ST:realdata}.

\begin{figure*}[htbp]\centering
\scalebox{0.80}{
\includegraphics[width=1.0\linewidth, trim=0 6.5pt 0 0, clip]{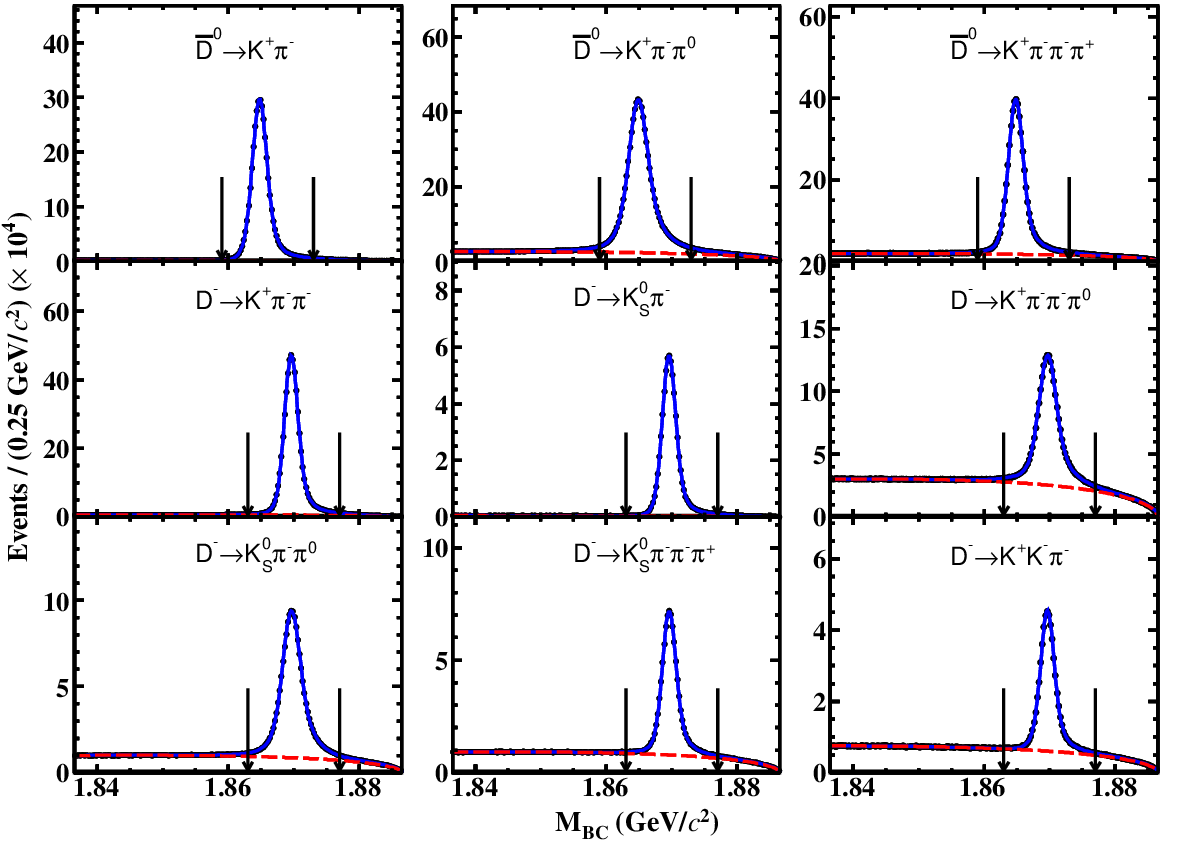}
}
\caption{
Fits to the $M_{\rm BC}$ distributions of the ST $\bar D$ candidates. In each plot, the points with error bars correspond to the data, the blue curves are the best fits, and 
the dashed red curves describe the fitted combinatorial background shapes.
The pair of black arrows indicate the $M_{\rm BC}$ signal window.}\label{fig:datafit_Massbc}
\end{figure*}

The ST yields are extracted by performing unbinned maximum likelihood fits to the corresponding $M_{\rm BC}$ distribution.
In the fit, the signal shape is derived from the MC-simulated signal shape convolved with a double-Gaussian function to compensate for the resolution difference between 
data and MC simulation.
The background shape is described by the ARGUS function~\cite{argus}, with the endpoint parameter fixed at 1.8865~GeV/$c^{2}$ corresponding to $E_{\rm beam}$.
Figure~\ref{fig:datafit_Massbc} shows the fits to the $M_{\rm BC}$ distributions of the accepted ST candidates in data for different ST modes.
The candidates with $M_{\rm BC}$ within $(1.859,1.873)$ GeV/$c^2$ for $\bar D^0$ tags and $(1.863,1.877)$ GeV/$c^2$ for $D^-$ tags are kept for further analyses.
Summing over the tag modes gives the total yields of ST $\bar D^0$ and $D^-$ mesons to be $(1613.5 \pm 0.5)\times 10^4$ and $(1064.7\pm0.4)\times 10^4$, where the 
uncertainties are statistical only.

\begin{table}
\renewcommand{\arraystretch}{1.2}
\centering
\caption {The $\Delta E$ requirements, the measured ST $\bar D$ yields in the data~($n^i_{\rm tag}$), and the ST efficiencies ($\varepsilon_{\rm tag}^{i}$) for nine tag modes. 
The uncertainties are statistical only.}
\scalebox{0.87}{
\begin{tabular}{lccc}
\hline
\hline
Tag mode & $\Delta E$~(GeV)  &  $n^i_{\rm tag}~(\times 10^3)$  &  $\varepsilon_{\rm tag}^{i}~(\%)    $       \\\hline
$\bar D^0\to K^+\pi^-$                    &  $(-0.027,0.027)$ & $3725.7\pm2.0$&$65.10\pm0.01$\\
$\bar D^0\to K^+\pi^-\pi^0$             &  $(-0.062,0.049)$ & $7422.3\pm3.2$&$35.60\pm0.00$\\
$\bar D^0\to K^+\pi^-\pi^-\pi^+$       &  $(-0.026,0.024)$ & $4987.5\pm2.5$&$40.94\pm0.01$\\
\hline
$D^-\to K^+\pi^-\pi^-$                   &  $(-0.025,0.024)$ & $5552.8\pm2.5$&$51.10\pm0.00$\\
$D^-\to K^{0}_{S}\pi^{-}$                 &  $(-0.025,0.026)$ & $\phantom{0}656.5\pm0.8$&$51.42\pm0.01$\\
$D^-\to K^{+}\pi^{-}\pi^{-}\pi^{0}$     &  $(-0.057,0.046)$ & $1723.7\pm1.8$&$24.40\pm0.00$\\
$D^-\to K^{0}_{S}\pi^{-}\pi^{0}$         &  $(-0.062,0.049)$ & $1442.4\pm1.5$&$26.45\pm0.00$\\
$D^-\to K^{0}_{S}\pi^{-}\pi^{-}\pi^{+}$ &  $(-0.028,0.027)$ & $\phantom{0}790.2\pm1.1$&$29.59\pm0.01$\\
$D^-\to K^{+}K^{-}\pi^{-}$                &  $(-0.024,0.023)$ & $\phantom{0}481.4\pm0.9$&$40.91\pm0.01$\\
\hline
\hline
          \end{tabular}
          }
          \label{ST:realdata}
          \end{table}

\begin{table*}
\centering
\caption {The $\Delta E_{\rm sig}$ requirements, the $M^{\rm sig}_{\rm BC}$ signal regions, the numbers of observed events $n_{\rm obs}$, and the estimated background yields 
$n_{\rm bkg1}^{\rm SB}$ and $n_{\rm bkg2}^{\rm MC}$ $\pm$ $\sigma_{\rm bkg2}^{\rm MC}$ in the $D^+$ and $D^0$ signal modes.}

\begin{tabular}{l @{\hspace{10pt}} c @{\hspace{10pt}} c @{\hspace{10pt}} c @{\hspace{10pt}} c @{\hspace{10pt}} c@{$\,\pm\,$}l}
\hline
\hline
Decays & $\Delta E_{\rm sig}$ (GeV) &  $M_{\rm BC}^{\rm sig}$ (GeV/$c^2$) &  $n_{\rm obs}$  &  $n_{\rm bkg1}^{\rm SB}$ & \multicolumn{2}{c}{$n_{\rm bkg2}^{\rm MC} \pm 
\sigma_{\rm bkg2}^{\rm MC}$}
\\\hline
$D^+\to \pi^+ \pi^0 e^+ e^-$ & $(-0.060,0.030)$ & $(1.864,1.877)$ & 11& 2 & 10.9 & 2.0 \\
$D^+\to K^+ \pi^0 e^+ e^-$   & $(-0.063,0.037)$ & $(1.862,1.877)$ & 0 & 0 & 1.5 & 0.4 \\
$D^+\to \pi^+ K_S^0 e^+ e^-$ & $(-0.038,0.020)$ & $(1.865,1.877)$ & 4 & 0 & 3.5 & 1.0 \\
$D^+\to K^+ K_S^0 e^+ e^-$   & $(-0.038,0.021)$ & $(1.865,1.875)$ & 0 & 0 & 0.2 & 0.1 \\
$D^+\to \rho^+ e^+ e^-$      & $(-0.060,0.030)$ & $(1.864,1.877)$ & 4 & 0 & 4.3 & 1.3 \\
$D^+\to K^{*+}_{K^+ \pi^0} e^+ e^-$    & $(-0.063,0.037)$ & $(1.862,1.877)$ & 0 & 0 & 0.4 & 0.1 \\
$D^+\to K^{*+}_{\pi^+ K_S^0} e^+ e^-$  & $(-0.038,0.020)$ & $(1.865,1.877)$ & 1 & 0 & 1.2 & 0.7\\
\hline
$D^0\to \pi^0 e^+ e^-$       & $(-0.094,0.031)$ & $(1.853,1.877)$ & 3 & 0 & 2.5 & 0.4\\
$D^0\to \eta e^+ e^-$        & $(-0.086,0.035)$ & $(1.854,1.878)$ & 1 & 0 & 1.8 & 0.3\\
$D^0\to K_S^0 e^+ e^-$       & $(-0.043,0.020)$ & $(1.858,1.873)$ & 4 & 0 & 1.6 & 0.3\\
$D^0\to \omega e^+ e^-$      & $(-0.076,0.035)$ & $(1.854,1.878)$ & 4 & 1 & 3.4 & 0.6\\
$D^0\to K_S^0 K_S^0 e^+ e^-$ & $(-0.056,0.041)$ & $(1.858,1.873)$ & 0 & 0 & 0.1 & 0.1\\
$D^0\to \pi^0 \pi^0 e^+ e^-$ & $(-0.113,0.067)$ & $(1.853,1.879)$ & 8& 0 & 14.6 & 6.0\\
$D^0\to \phi e^+ e^-$        & $(-0.044,0.015)$ & $(1.858,1.872)$ & 0 & 0 & 0.1 & 0.1\\
$D^0\to \rho^0 e^+ e^-$      & $(-0.053,0.020)$ & $(1.857,1.873)$ & 4 & 0 & 4.1 & 0.4\\
$D^0\to \eta^{\prime}_{\gamma \pi^+ \pi^-}e^+e^-$ & $(-0.050,0.039)$ &$(1.854,1.878)$ & 6 & 0 & 5.4 & 1.8\\
$D^0\to \eta^{\prime}_{\eta \pi^+ \pi^-}e^+e^-$ & $(-0.067,0.044)$ & $(1.854,1.878)$  & 0 & 0 & 0.2 & 0.1\\
\hline
\hline
\end{tabular}
\label{DT}
\end{table*}

\section{Signal selection}
Signal candidates of the 15 rare decays of $D\to h(h^{(')})e^{+}e^{-}$  are reconstructed in the remaining charged tracks and showers recoiling against the ST $D$ mesons.
The selection criteria for the charged tracks and neutral showers are the same as those used in the ST event selection.
Positrons and electrons are identified using a combined likelihood \(\mathcal{L}'\) constructed from MDC, TOF, and EMC information. The likelihood is evaluated under electron, pion, and kaon hypotheses. An electron (positron) candidate is required to satisfy $\mathcal{L}'(e) > 0.001$ and $\mathcal{L}'(e) > \mathcal{L}'(e)/(\mathcal{L}'(e) + \mathcal{L}'(\pi) + \mathcal{L}'(K)) > 0.8$.  In addition, for the higher-momentum track in the pair, the ratio of energy deposited in the EMC to momentum measured in the MDC, $E/p$, must satisfy $(0.8 < E/p < 1.2$. 
Studies of inclusive MC samples show that the selected $e^{+}e^{-}$ pairs predominantly arise from $\gamma$-conversions, with the photons originating from decays of intermediate states and the conversions occur around the beam pipe and the inner wall of the MDC.
To suppress these backgrounds, the $e^+e^-$ pair is required to originate from a vertex reconstructed outside a cylindrical region of radius $R_{xy}\in (2.0, 8.0)~\text{cm}$ in the transverse plane.

For the $K_S^0$ candidate, in addition to the same criteria used in the ST event selection, the candidate with the smallest $\chi^2$ is retained under the requirement $\chi^2 
< 20$.
The $\eta$ and $\pi^0$ meson candidates are reconstructed via their $\gamma\gamma$ decay modes, requiring the invariant mass $M_{\gamma\gamma}$ to be within (0.505, 0.570) 
GeV/$c^2$ and (0.115, 0.150) GeV/$c^2$.
The two-pion decay modes are used to reconstruct $\rho^0$ and $\rho^+$ candidates, with requirements on the invariant mass set to the range (0.625, 0.925) GeV/$c^2$ and (0.64, 
0.89) GeV/$c^2$, respectively.
Similarly, the $K^+K^-$ and $\pi^+\pi^-\pi^0$ decay modes are used to reconstruct $\phi$ and $\omega$ candidates, with requirements on the invariant mass set to (1.01, 1.03) 
GeV/$c^2$ and (0.720, 0.840) GeV/$c^2$. Candidates for the $K^{*+}$ and $\eta'$ mesons are reconstructed via two decay modes each.
For the $K^{*+}$, the reconstruction requires the invariant mass of the $K^+\pi^0$ or $\pi^+K_S^0$ to fall within (0.8, 1.0) GeV/$c^2$.
For the $\eta'$, the invariant masses of the $\gamma\pi^+\pi^-$ and $\eta\pi^+\pi^-$ are required to be in the ranges of (0.939, 0.974) GeV/$c^2$ or (0.943, 0.972) GeV/$c^2$, 
respectively.

Similar to the ST selection, $\Delta E$ and $M_{\rm BC}$ for the signal candidates of the rare $D$ decays in DT events, denoted as $\Delta E_{\rm sig}$ and $M_{\rm BC}^{\rm 
sig}$, are calculated.
For each signal mode, $\Delta E_{\rm sig}$ is required to be within 3$\sigma$ of the nominal value, as listed in Table~\ref{DT}, and only the combination with the smallest 
$|\Delta E_{\rm sig}|$ is kept.
A dominant background arises from $\pi^0 \to \gamma e^+e^-$ and $\eta \to \gamma e^+e^-$ decays, where the photon escapes reconstruction.  
For $\pi^0$ decays, this background is suppressed by requiring the invariant mass of the electron-positron pair $M_{e^+e^-} > 0.2~\text{GeV}/c^2$, which reduces the contribution to below 1\% of its original level.  
For $\eta$ decays, the background is vetoed by reconstructing the missing photon from combinations of the $e^+e^-$ pair with any unused photon in the event. The combination yielding an invariant mass closest to the nominal $\eta$ mass is retained; events falling within a $\pm 2\sigma$ window of $(0.505, 0.570)~\text{GeV}/c^2$ around the $\eta$ peak are rejected. This veto removes over 50\% of the remaining background.
Furthermore, to veto the contribution from $D \to h(h^{(')})\phi$, $\phi \to e^{+}e^{-}$ decays, $M_{e^{+}e^{-}}$ is required to be outside of the $\phi$ mass region, defined 
as (0.935, 1.053) GeV/$c^2$.

After applying all the aforementioned selection criteria, the backgrounds are separated into two categories: events with an incorrectly reconstructed ST candidate, and events 
with a correctly reconstructed ST candidate but a wrong signal candidate.
The former background can be estimated with the surviving events in the ST sideband (SB) region of $M_{\rm BC}^{\rm tag}$ distribution, which is defined as (1.830, 1.855) 
GeV/$c^2$ for $\bar D^0$ decays and (1.830, 1.860) GeV/$c^2$ for $D^-$ decays.
The latter predominantly originate from the process where $D \to h(h^{(\prime)}) \pi^+ \pi^-$ events are mis-reconstructed as $D \to h(h^{(\prime)}) e^+ e^-$ and from the $D 
\to h(h^{(\prime)}) \eta, \eta \to {\gamma}e^{+}e^{-}$ process. These two contributions account for approximately 50\% and 35\% of the background, respectively.
The corresponding number of incorrectly reconstructed ST background events, $n_{\rm bkg1}$, is estimated with the number of events in the SB region $n_{\rm bkg1}^{\rm SB}$ 
normalized by a scale factor $f$, which is the ratio of the integrated numbers of background events in the signal and SB regions.
The scale factors $f$ are determined to be 0.466 $\pm$ 0.001 and 0.611 $\pm$ 0.001 for the charged and neutral $D$ decays~\cite{sys_hhee}, respectively, where the 
uncertainties are statistical only.
The incorrectly reconstructed ST background is expected to follow a Poisson $(\mathcal{P})$ distribution with central value of $n_{\rm bkg1}^{\rm SB}\cdot f$. The background 
from mis-reconstructed signal, $n_{\rm bkg2}$, is estimated with the $D^{+} D^{-}$ and $D^{0}\bar D^{0}$ events in the inclusive MC samples by subtracting the incorrectly 
reconstructed ST events, and the corresponding number of events is expected to follow a Gaussian distribution $(\mathcal{G})$.
This Gaussian distribution $\mathcal{G}$ is centered at $n_{\rm bkg2}^{\rm MC}$ with an uncertainty of $\sigma_{\rm bkg2}^{\rm MC}$, where the uncertainty $\sigma_{\rm 
bkg2}^{\rm MC}$ includes both statistical uncertainty and the uncertainty from BF of the dominant backgrounds from decays $D \to h(h^{(\prime)}) \pi^{+}\pi^{-}$ and $D \to 
h(h^{(\prime)}) \eta, \eta \to {\gamma}e^{+}e^{-}$.

\begin{figure*}[htbp]
\includegraphics[width=1.0\linewidth, trim=0 10pt 0 0, clip]{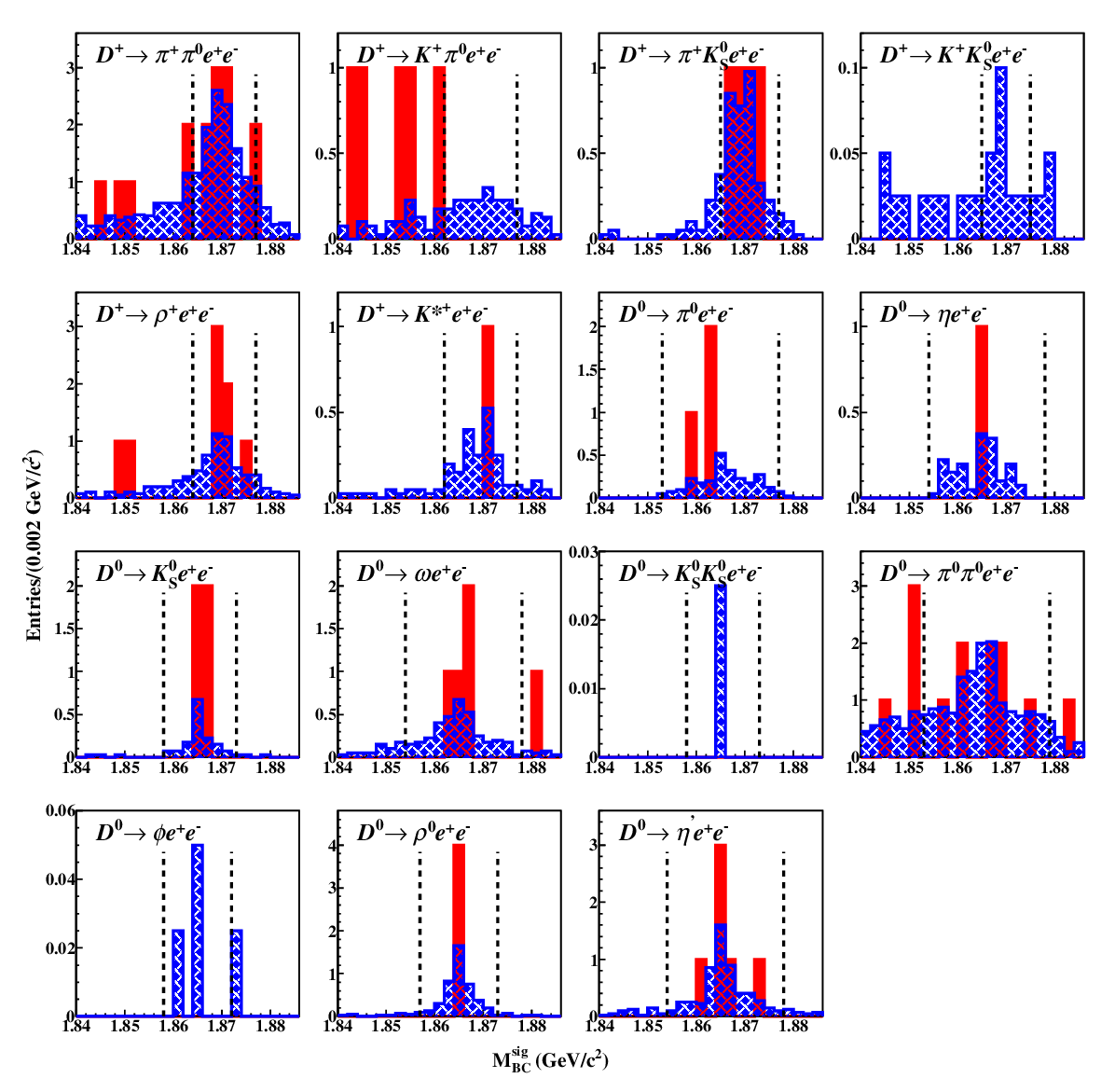}
\caption{Distributions of $M_{\rm BC}^{\rm sig}$ for the signal modes after applying all selection criteria. The solid histograms are data, the hatched ones are the background 
events in the inclusive MC samples scaled to the luminosity of data. The black dashed lines denote the signal region.
 }
\label{mbc}
\end{figure*}

The number of observed events $n_{\rm obs}$ and the estimated background yields $n_{\rm bkg1,2}$ are listed in Table~\ref{DT}.
The detection efficiencies for the DT $D^0$ and $D^+$ decays are provided in Table~\ref{eff_Dn} and Table~\ref{eff_Dc}, respectively.
As shown in Fig.~\ref{mbc}, no significant excess over the expected backgrounds is observed.
The ULs of the BFs can be determined by subtracting the background contributions from the observed signal candidates $n_{\rm obs}$ and correcting the reconstructed 
efficiencies.
The systematic uncertainties, which are described in the following, are also accounted for.

\begin{table}[htbp]
      \centering
       \caption{MC-determined DT detection efficiencies of the different $D^0$ decay modes (\%). The uncertainties are all statistical.}
     \begin{tabular}{l r@{$\,\pm\,$}l r@{$\,\pm\,$}l r@{$\,\pm\,$}l}
  \hline
  \hline
  \multicolumn{1}{c}{Signal vs.} & \multicolumn{2}{c}{$K^+\pi^-$} & \multicolumn{2}{c}{$K^+\pi^-\pi^0$} & \multicolumn{2}{c}{$K^+\pi^-\pi^-\pi^+$} \\
  \hline
$\pi^0 e^+ e^-$                 & 26.94 & 0.08 & 14.25 & 0.04 & 15.54 & 0.04 \\
$\eta e^+ e^-$                  & 23.30 & 0.07 & 12.48 & 0.03 & 13.37 & 0.03 \\
$K_S^0 e^+ e^-$                 & 14.98 & 0.04 & 7.95 & 0.02 & 8.65 & 0.02 \\
$\omega e^+ e^-$                & 13.99 & 0.03 & 6.80 & 0.02 & 7.45 & 0.02 \\
$K^0_S K^0_S e^+ e^-$           &  2.71 & 0.01 & 1.37 & 0.01 & 1.35 & 0.01 \\
$\pi^0 \pi^0 e^+ e^-$           & 12.85 & 0.04 & 6.12 & 0.03 & 7.02 & 0.03 \\
$\phi e^+ e^-$                  &  4.14 & 0.02 & 2.24 & 0.01 & 2.27 & 0.01 \\
$\rho^0 e^+ e^-$                & 24.34 & 0.08 & 12.60 & 0.04 & 13.81 & 0.04 \\
$\eta^{\prime} e^+ e^-$         & 13.44 & 0.05 & 6.57 & 0.03 & 6.89 & 0.03 \\
  \hline
  \hline
\end{tabular}
             \label{eff_Dn}
\end{table}

\begin{table*}[htp]
      \centering
       \caption{MC-determined DT detection efficiencies of the different $D^+$ decay modes (\%). The uncertainties are all statistical.}
      \begin{tabular}{l r@{$\,\pm\,$}l @{\hspace{10pt}} r@{$\,\pm\,$}l @{\hspace{10pt}} r@{$\,\pm\,$}l @{\hspace{10pt}} r@{$\,\pm\,$}l @{\hspace{10pt}} r@{$\,\pm\,$}l 
      @{\hspace{10pt}} r@{$\,\pm\,$}l}
    \hline
    \hline
    \multicolumn{1}{c}{Signal vs.} & \multicolumn{2}{c}{$K^+\pi^-\pi^-$} & \multicolumn{2}{c}{$K^0_S\pi^-$} & \multicolumn{2}{c}{$K^+\pi^-\pi^-\pi^0$} & 
    \multicolumn{2}{c}{$K^0_S\pi^-\pi^0$} & \multicolumn{2}{c}{$K^0_S\pi^-\pi^-\pi^+$} & \multicolumn{2}{c}{$K^+K^-\pi^-$} \\
    \hline
    $\pi^+\pi^0e^+e^-$ & 12.75 & 0.03 & 14.02 & 0.04 & 5.34 & 0.02 & 6.33 & 0.02 & 7.45 & 0.03 & 10.52 & 0.03 \\
    $K^+\pi^0e^+e^-$ & 10.12 & 0.03 & 10.96 & 0.03 & 3.84 & 0.01 & 5.17 & 0.02 & 6.01 & 0.02 & 7.89 & 0.02 \\
    $\pi^+K^0_Se^+e^-$ & 7.74 & 0.03 & 8.24 & 0.03 & 3.17 & 0.01 & 3.66 & 0.02 & 3.23 & 0.01 & 6.02 & 0.02 \\
    $K^+K^0_Se^+e^-$ & 3.18 & 0.01 & 3.39 & 0.01 & 1.22 & 0.01 & 1.48 & 0.01 & 1.37 & 0.01 & 2.43 & 0.01 \\
    $\rho^+e^+e^-$ & 10.56 & 0.04 & 12.27 & 0.04 & 4.40 & 0.02 & 5.19 & 0.02 & 6.05 & 0.02 & 8.19 & 0.04 \\
    $K^{*+}e^+e^-$ & 5.94 & 0.04 & 7.65 & 0.04 & 3.00 & 0.02 & 3.50 & 0.04 & 3.30 & 0.02 & 4.41 & 0.02 \\
    \hline
    \hline
  \end{tabular}
             \label{eff_Dc}
\end{table*}

\section{Systematic uncertainty}
With the DT technique, the systematic uncertainties in the BF measurements due to the detection and reconstruction of the ST $D$ mesons mostly cancel, as shown in 
Eq.~(\ref{br}).
The uncertainty associated with the ST yield $N_{\rm ST}^{\rm tot}$, is assigned as 0.3\% after varying the signal, background shapes and floating the parameters of one 
Gaussian in the fit.

For the signal side, the following sources of systematic uncertainties, as summarized in Table~\ref{sys1} and Table~\ref{sys2}, are considered.
All of these contributions are added in quadrature to obtain the total systematic uncertainties. The uncertainties of tracking and PID for $K^{\pm}$ and $\pi^{\pm}$ mesons are 
studied with control samples of $D \bar{D}$ favored hadronic decay modes~\cite{kpi_trk_pid}.
We assign an uncertainty of 0.3\% per track for the tracking and 0.3\% for the PID uncertainties.
The tracking, PID and $E/p$ efficiency for electron or positron detection is studied using radiative Bhabha events. The corresponding systematic uncertainty, evaluated by 
weighting according to the $\rm{\cos}\theta$ and transverse momentum distributions of the $e^{\pm}$ tracks, ranges from 0.5\% to 3.4\% for the various decay processes.
The uncertainties for vetoes of $\pi^0 \to \gamma e^{+} e^{-}$ decays and $\eta \to \gamma e^{+} e^{-}$ decays are also studied by using the control sample of radiative Bhabha events.
The $e^{\pm}$ momentum resolution is determined from both data and MC in bins of momentum, and the relative difference between them is taken as a systematic uncertainty and assigned to the signal efficiency.
A 2.0\% systematic uncertainty is assigned for $\pi^0$ meson reconstruction, based on control-sample studies of $D^0 \to K^- \pi^+ \pi^0$ decays that compare the reconstruction efficiency between data and MC.
According to the momentum dependent data-MC difference, the uncertainties of $\eta$, $\omega$, and $K_S^0$ meson reconstruction are studied using control samples of $D\bar{D}$ 
events, with uncertainties of 1.5\% for $K_S^0$, 0.8\% for $\omega$, and 1.2\% for $\eta$~\cite{sys_hhee}.
Similarly, an uncertainty of 1.1\% for $\rho^0$ and $\rho^+$~\cite{sys_rho} mesons, 0.5\% for $\phi$ and $\eta^{\prime}$~\cite{sys_phi_etap} mesons and 0.1\% for the 
$K^{*+}$~\cite{sys_kstar} meson are assigned.

The $\gamma$-conversion background is suppressed by a requirement on the distance from the reconstructed vertex of the $e^{+}e^{-}$ pair to the IP. The uncertainty due to this 
requirement is studied using a sample of $J/\psi \to \pi^+ \pi^- \pi^0$ decays with the $\pi^0$ subsequently decaying to the $ \gamma e^+ e^-$ final state~\cite{ee2}. The 
relative difference of the efficiency between data and MC simulation is 1.8\%, and is assigned as the uncertainty.

The estimated signal detection efficiencies are derived from MC simulations and computed as a weighted average of the efficiencies from the LD and SD samples.
The weighting is based on the theoretically predicted BFs of the LD and SD processes~\cite{hhll}.
The difference between the signal and LD efficiencies is assigned as a systematic uncertainty.
The uncertainties from the BFs of the intermediate states decays of the neutral mesons, $\mathcal{B}_{\rm inter}$, are assigned according to the PDG values~\cite{pdg2025}.

\begin{table*}[htp]
      \centering
       \caption{Relative systematic uncertainties on the BFs of different $D^0$ decay modes (\%).}

      \begin{tabular}{l c @{\hspace{5pt}} c @{\hspace{5pt}} c @{\hspace{5pt}} c @{\hspace{5pt}} c @{\hspace{5pt}} c @{\hspace{5pt}} c @{\hspace{5pt}}c @{\hspace{5pt}} c  }
  \hline
  \hline
  Source   &$\pi^{0} e^+e^-$ &  $\eta e^+e^-$& $\omega e^+e^-$& $K_S^{0} e^+e^-$&$\phi e^+e^-$   &$\rho^0 e^+e^-$ &  $K_S^{0} K_S^{0} e^+e^-$& $\pi^{0} \pi^{0} e^+e^-$& 
  $\eta^{\prime} e^+e^-$\\
  \hline
  $N_{\rm{ST}}^{\rm tot}$ &0.3 &0.3 &0.3 &0.3&0.3 &0.3  &0.3 &0.3 &0.3   \\
$K^{\pm}/\pi^{\pm}$ tracking  & - & - & 0.6 & - & 0.6 & - & - & - & 0.6  \\
$K^{\pm}/\pi^{\pm}$ PID  & - & - & 0.6 & - & 0.6 & - & - & - & 0.6   \\
$e^{\pm}$ tracking, PID and $E/p$&  0.5 & 0.7 & 1.5 & 0.8 & 2.0 & 1.3  & 3.4 & 0.6 & 1.9  \\
$\pi^{0}$/$\eta$ reconstruction  & 2.0 & 1.2 & 2.0 & -  & - & -  & - & 4.0 & -  \\
$K_S^{0}$/$\omega$/$\rho^0$/$\eta^{\prime}$ reconstruction  & - & - & 0.8 & 1.5 & 0.2 & 1.1 & 3.0 & - & 0.5  \\
Veto of $\gamma$ conversion & 1.8 & 1.8  & 1.8 & 1.8& 1.8 & 1.8 & 1.8 & 1.8 & 1.8   \\
MC modeling &  1.0 & 2.1& 2.3& 2.7&  5.4 & 3.5& 2.9& 3.6& 2.8 \\
Veto of $\pi^{0}\to\gamma e^{+}e^{-}$  & 1.1 & 1.3 & 1.3 & 1.3 & 1.7 & 1.3 & 1.3 & 1.3 & 1.3   \\
 Veto of $\eta\to\gamma e^{+}e^{-}$ & 1.1 & 1.3 & 1.3 & 1.3 & 1.7 & 1.3 & 1.3 & 1.3 & 1.3  \\
$\mathcal{B}_{\rm inter}$  & 0.1 & 0.5 & 0.8 & 0.1 & 1.0 & 0.1 & 0.1 & 0.1 & 1.2  \\
\hline
Total & 3.3 &3.7 &4.5 &4.1 &6.6 &4.7 & 6.0 &5.9 &4.4   \\
                   \hline
                   \hline
\end{tabular}
             \label{sys1}
            \end{table*}

\begin{table*}[htp]
      \centering
       \caption{Relative systematic uncertainties on the BFs of different $D^+$ decay modes (\%).}
      \begin{tabular}{l c @{\hspace{10pt}}c @{\hspace{10pt}}c @{\hspace{10pt}}c @{\hspace{10pt}}c @{\hspace{10pt}}c }
  \hline
  \hline
  Source  &$\pi^+ \pi^{0} e^+e^-$& $K^+ \pi^{0} e^+e^-$& $\pi^+ K_S^{0}e^+e^-$& $K^+ K_S^{0}e^+e^-$ & $\rho^{+} e^+e^-$& $K^{*+} e^+e^-$  \\
  \hline
  $N_{\rm{ST}}^{\rm tot}$&0.3 &0.3 &0.3 &0.3&0.3 &0.3  \\
$K^{\pm}/\pi^{\pm}$ tracking  & 0.3 & 0.3 & 0.3 & 0.3 & - & 0.3  \\
$K^{\pm}/\pi^{\pm}$ PID  & 0.3 & 0.3 & 0.3 & 0.3 & - & 0.3  \\
$e^{\pm}$ tracking, PID and $E/p$&  1.9 & 2.3 & 2.2 & 3.3&  1.7 & 2.3\\
$\pi^{0}$ reconstruction & 2.0 & 2.0 & - & -  & - & 2.0\\
$K_S^{0}$/$\rho^+$/$K^{*+}$ reconstruction  & - & - & 1.5 & 1.5& 1.1 & 0.1\\
Veto of $\gamma$ conversion  & 1.8 & 1.8 & 1.8 & 1.8& 1.8 & 1.8 \\
MC modeling & 3.3&2.2 &3.5 & 0.9&3.4&2.9 \\
Veto of $\pi^{0}\to\gamma e^{+}e^{-}$  & 1.5 & 1.6 & 1.6 & 1.7& 1.5 & 1.6 \\
 Veto of $\eta\to\gamma e^{+}e^{-}$ & 1.5 & 1.6 & 1.6 & 1.7& 1.5 & 1.6\\
$\mathcal{B}_{\rm inter}$  & 0.1 & 0.1 & 0.1 & 0.1& 0.1 & 0.1\\
\hline
Total  &5.1 &4.8 &5.1 &4.9 &5.2 &5.1 \\
                   \hline
                   \hline
\end{tabular}
             \label{sys2}
            \end{table*}

\section{Results}
After applying all selection criteria to the data, no significant signals are observed. The signal yields are consistent with the background estimates obtained from MC 
simulations.
To calculate the ULs on the BFs for the signal decays, we use a maximum likelihood estimator, extended from the Profile Likelihood method~\cite{plh}. The
number of observed events $n_{\rm obs}$ and the number of background events estimated with the sideband $n_{\mathrm{bkg}2}$ are expected to follow a Poisson distribution. The 
detection efficiency and the number of background events estimated with MC simulation $n_{\mathrm{bkg}1}$ are expected to follow a Gaussian distribution.
The joint likelihood function is expressed as
\begin{equation}
    \label{br1}
   {
\begin{aligned}
&\mathcal{L}{(n_{\rm bkg1}, n_{\rm bkg2}, \varepsilon_{\mathrm{sig~}}, \mathcal{B})}=\\
&\mathcal{P}(n_{\mathrm{obs}}| n_{\mathrm{tag}}\cdot\mathcal{B}\cdot\varepsilon_{\mathrm{sig}}+n_{\mathrm{bkg}1}+n_{\mathrm{bkg}2}) \\
&\times\mathcal{G}(\varepsilon_{\mathrm{sig}}^{\mathrm{MC}}|\varepsilon_{\mathrm{sig}},\varepsilon_{\mathrm{sig}}^{\mathrm{MC}}\cdot\sigma_{\varepsilon}^{\mathrm{MC}}) \\
&\times\mathcal{P}(n_{\mathrm{bkg}1}^{\mathrm{SB}}\cdot 
f|n_{\mathrm{bkg}1})\times\mathcal{G}(n_{\mathrm{bkg}2}^{\mathrm{MC}}|n_{\mathrm{bkg}2},\sigma_{\mathrm{bkg}2}^{\mathrm{MC}}).
\end{aligned}}
    \end{equation}

\begin{figure*}[htbp]
\includegraphics[width=1.0\linewidth, trim=0 7pt 0 0, clip]{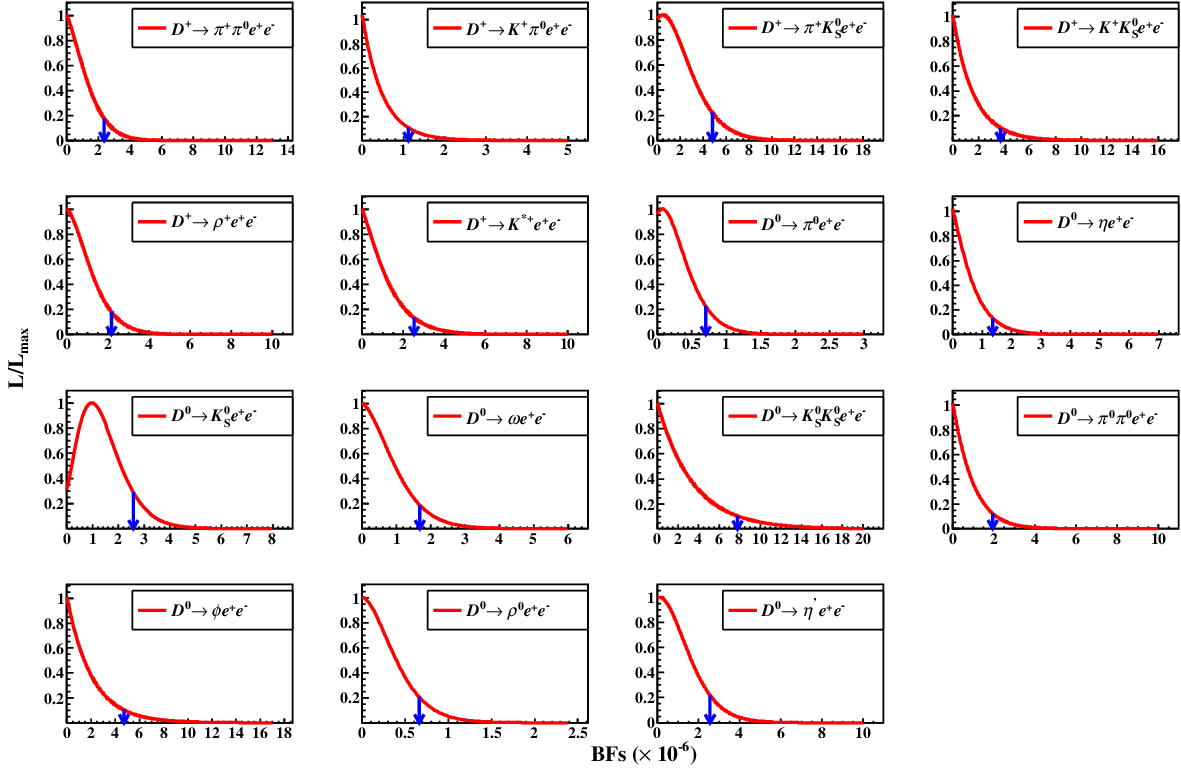}
\caption{Likelihood curves as a function of the signal BFs. The arrows point to the position of the ULs at the 90\%C.L..
 }
\label{ULs}
\end{figure*}

Here $\mathcal{B}$ denotes the BF of the signal decay and
$n_{\mathrm{tag}}$ is the total number of ST events $n_{\mathrm{tag}}$ = $\sum_{i}n_{\mathrm{tag}}^{i}$. The parameter
$\varepsilon_{\mathrm{sig~}}^{\rm MC}$ is therefore the averaged signal efficiency over different ST modes.
It can be calculated as $\varepsilon_{\mathrm{sig~}}^{\rm MC} =(\sum_{i}n_{\mathrm{tag~}}^{i}\cdot\varepsilon_{\mathrm{tag,sig}}^{i,\rm MC}/\varepsilon_{\mathrm{tag}}^{i,\rm 
MC})/n_{\mathrm{tag}}$ according to Eq.~(\ref{br}).
The relative statistical uncertainties listed in Tables~{\ref{eff_Dn} and \ref{eff_Dc}}, and the systematic uncertainties as given in Tables{~\ref{sys1} and \ref{sys2}} are 
included through $\sigma_{\varepsilon}^{\rm MC}$. The likelihood function $\mathcal{L}$ is maximized in the parameter space $\theta_{\alpha}$ = ($n_{\rm bkg1}$, $n_{\rm 
bkg2}$, $\varepsilon_{\mathrm{sig~}}$, $\mathcal{B}$) to determine $\mathcal{L}_{\rm max}$.
For each value of $\mathcal{B}$, the likelihood function can also be maximized in the space of the nuisance parameters ($n_{\rm bkg1}$, $n_{\rm bkg2}$, 
$\varepsilon_{\mathrm{sig~}}$) to determine $\mathcal{L}$.
The resultant curves of the $\mathcal{L}/\mathcal{L}_{\rm max}$ versus the $\mathcal{B}$ for all signal modes are shown in Fig.~\ref{ULs}. The ULs on the signal BFs at the 
90\% CL are estimated by integrating the likelihood curves in the physical region of $\mathcal{B} \geq 0$, as listed in Table~\ref{results}.
We have also measured the ULs on the BFs without including systematic uncertainties. The difference between the ULs obtained with and without systematic uncertainties is less 
than 0.1\%.
This analysis improves the ULs on the BFs for eight decay channels listed in the PDG, which are primarily based on BESIII~\cite{sys_hhee} and CLEO measurements.
This analysis updates the upper limits on the branching fractions for ten decay channels listed in the PDG, of which eight are primarily based on previous measurements from BESIII~\cite{sys_hhee}. 
Compared to previous measurements, the dataset used in this work is larger and the selection is further optimized to suppress backgrounds and enhance detection efficiency.

 \begin{table*}
      \centering
       \caption{The ULs on the BFs of each signal decay obtained in this work, and the comparison with the PDG values. }
     \begin{tabular}{l @{\hspace{35pt}} c @{\hspace{35pt}} c }
  \hline
  \hline
  Process & $\mathcal{B} (\times10^{-6})$ &PDG~\cite{pdg2025}$(\times10^{-5})$ \\
  \hline
$D^+\to \pi^+ \pi^0 e^+ e^-$    & $<2.4  $        &  $<1.4 $ \\
$D^+\to K^+ \pi^0 e^+ e^-$      & $<1.1  $   & $<1.5 $\\
$D^+\to \pi^+ K_S^0 e^+ e^-$ & $<4.8  $&$<2.6 $ \\
$D^+\to K^+ K_S^0 e^+ e^-$  & $<3.7  $ & $<1.1 $ \\
$D^+\to \rho^{+}  e^+ e^-$   & $<2.1  $ &  - \\
$D^+\to K^{*+}  e^+ e^-$   & $<2.5$ & - \\
\hline

$D^0\to \pi^0 e^+ e^-$    & $<0.7  $     & $<0.4 $ \\

$D^0\to \eta e^+ e^-$    & $<1.4  $        & $<0.3 $\\

$D^0\to \omega e^+ e^-$      & $<1.6  $    &  $<0.8 $\\
$D^0\to K_S^0 e^+ e^-$   & $<2.6  $      & $<1.2 $\\
$D^0\to K_S^0 K_S^0 e^+ e^-$    & $<7.8  $&  -\\
$D^0\to \pi^0 \pi^0 e^+ e^-$ &$<1.9  $&  -\\
$D^0\to \phi  e^+ e^-$  &  $<4.6  $&  $<5.2 $ \\
$D^0\to \rho^0  e^+ e^-$ & $<0.7  $&  $<10.0$ \\
$D^0\to \eta^{\prime}  e^+ e^-$ &$<2.5  $& -  \\
                   \hline
                   \hline
\end{tabular}
             \label{results}
            \end{table*}

\section{Summary}
We search for 15 rare decays of $h(h^{(')})e^{+}e^{-}$ based on the DT analysis using 20.3 fb$^{-1}$ of $e^+ e^-$ collision data collected at the center-of-mass energy of 
3.773 GeV with the BESIII detector.
No significant signals are observed, and the corresponding ULs on the BFs are reported to be at $10^{-6}$ $\sim$ $10^{-7}$ level at the 90\% C.L..
In this work, the ULs on the BFs of $D^+\to \rho^{+}  e^+ e^-$, $D^+\to K^{*+}  e^+ e^-$, $D^0\to K_S^0 K_S^0 e^+ e^-$, $D^0\to \pi^0 \pi^0 e^+ e^-$ and $D^0\to \eta^{\prime}  
e^+ e^-$ decay channels are measured for the first time, which are between $2.1\times 10^{-6}$ and $7.8\times 10^{-6}$.
For the $D^0\to \rho^{0} e^+ e^-$ and $D^0\to \phi e^+ e^-$ decays, the ULs on the BFs are $0.7\times 10^{-6}$ and $4.6\times 10^{-6}$, respectively.
For the remaining eight decay channels, the ULs on the BFs have been improved by a factor of 4 $\sim$ 14 compared to the previous PDG values.
These results provide crucial input for more stringent tests of the SM predictions of lepton flavor universality.

\section{Acknowledgement}
The BESIII Collaboration thanks the staff of BEPCII (https://cstr.cn/31109.02.BEPC) and the IHEP computing center for their strong support. This work is supported in part by 
National Key R\&D Program of China under Contracts Nos. 2023YFA1606000, 2023YFA1606704; National Natural Science Foundation of China (NSFC) under Contracts Nos. 11635010, 
11935015, 11935016, 11935018, 12025502, 12035009, 12035013, 12061131003, 12192260, 12192261, 12192262, 12192263, 12192264, 12192265, 12221005, 12225509, 12235017, 12342502, 
12361141819; the Chinese Academy of Sciences (CAS) Large-Scale Scientific Facility Program; the Strategic Priority Research Program of Chinese Academy of Sciences under 
Contract No. XDA0480600; CAS under Contract No. YSBR-101; 100 Talents Program of CAS; The Institute of Nuclear and Particle Physics (INPAC) and Shanghai Key Laboratory for 
Particle Physics and Cosmology; ERC under Contract No. 758462; German Research Foundation DFG under Contract No. FOR5327; Istituto Nazionale di Fisica Nucleare, Italy; Knut 
and Alice Wallenberg Foundation under Contracts Nos. 2021.0174, 2021.0299, 2023.0315; Ministry of Development of Turkey under Contract No. DPT2006K-120470; National Research 
Foundation of Korea under Contract No. NRF-2022R1A2C1092335; National Science and Technology fund of Mongolia; Polish National Science Centre under Contract No. 
2024/53/B/ST2/00975; STFC (United Kingdom); Swedish Research Council under Contract No. 2019.04595; U. S. Department of Energy under Contract No. DE-FG02-05ER41374

\end{document}